\documentclass[a4paper,10pt,oneside]{article}

\usepackage{amssymb}
\usepackage{amsmath}
\usepackage{amsthm}

\makeatletter
\let\@afterindentfalse\@afterindenttrue
\@afterindenttrue
\makeatother

\newtheorem{theorem}{Theorem}

\newcommand{\ci}{\mathop{\mathrm{i}}\nolimits}
\newcommand{\Cov}{\mathop{\mathrm{Cov}}\nolimits}
\newcommand{\cpre}[1]{{}^{*}\hskip-3pt {#1}}
\newcommand{\dint}{\mathop{\mathrm{d}}\nolimits\hskip-1pt}
\newcommand{\Fop}{\mathcal{F}_{\mathrm{op}}}
\newcommand{\Fopn}{\mathcal{F}_{\mathrm{op}}^{(\mathrm{n})}}
\newcommand{\Fopr}{\mathcal{F}_{\mathrm{op}}^{(\mathrm{r})}}
\newcommand{\Ima}{\mathop{\mathrm{Im}}\nolimits}
\newcommand{\Mm}{\mathcal{M}_{n}}
\newcommand{\MN}{\mathcal{M}_{n}}
\newcommand{\MT}{M_{n,\mathrm{sa}}}
\newcommand{\Qov}{\mathop{\mathrm{Qov}}\nolimits}
\newcommand{\Rea}{\mathop{\mathrm{Re}}\nolimits}
\newcommand{\TMN}{M_{n,\mathrm{sa}}^{(0)}}
\newcommand{\Tr}{\mathop{\mathrm{Tr}}\nolimits}
\newcommand{\Var}{\mathop{\mathrm{Var}}\nolimits}

\begin{document}

\title{
Uncertainty principle with quantum Fisher information
  \thanks{keywords: uncertainty principle, quantum Fisher information;
          MSC: 62B10, 94A17}}
\author{Attila Andai\thanks{andaia@math.bme.hu}\\
  RIKEN, BSI, Amari Research Unit \\
  2--1, Hirosawa, Wako, Saitama 351-0198, Japan.}
\date{October 10, 2007}

\maketitle

\begin{abstract}
In this paper we prove a nontrivial lower bound for the determinant of the covariance matrix of quantum
  mechanical observables, which was conjectured by Gibilisco, Isola and Imparato.
The lower bound is given in terms of the commutator of the state and the observables and their scalar
  product, which is generated by an arbitrary symmetric operator monotone function.
\end{abstract}

\section*{Introduction}

The basic object in the statistical description of a classical physical system is a probability space
  $(\Omega,\mathcal{B},\mu)$, where the measure $\mu$ determines the state of the system and the physical
  quantities are measurable $\Omega\to\mathbb{R}$ functions.
The covariance of the quantities $X,Y\in L^{2}(\Omega,\mathbb{R},\mu)$ is defined as
\begin{equation*}
\Cov_{\mu}(X,Y)=\int_{\Omega}XY\dint\mu-\left(\int_{\Omega}X\dint\mu\right)
  \left(\int_{\Omega}Y\dint\mu\right)
\end{equation*}
  and the variance of a quantity is $\Var_{\mu}(X)=\Cov_{\mu}(X,X)$.
The Cauchy--Schwartz inequality in this setting gives
\begin{equation*}
\Var_{\mu}(X)\Var_{\mu}(Y)-\Cov_{\mu}(X,Y)^{2}\geq 0
\end{equation*}
  which can be reformulated as
\begin{equation*}
\det\begin{pmatrix}
\Cov_{\mu}(X,X) & \Cov_{\mu}(X,Y)\\
\Cov_{\mu}(Y,X) & \Cov_{\mu}(Y,Y)\end{pmatrix}\geq 0.
\end{equation*}

The quantum mechanical Hilbert space formalism gives a mathematical description of particles with
  spin of $\frac{n-1}{2}$.
Concentrating on the spin part of non relativistic particles one can build a proper mathematical model
  in an $n$ dimensional complex Hilbert space.
This is the simplest physical realization of an $n$-level quantum system.
The states of an $n$-level system are identified with the set of positive semidefinite self-adjoint
  $n\times n$ matrices of trace 1, and the physical observables are identified with the set of
  self-adjoint $n\times n$ matrices.
For a given state $D$ the (symmetrized) covariance of the observables $A$ and $B$ is defined as
\begin{equation*}
\Cov_{D}(A,B)=\frac{1}{2}\Bigl(\Tr(DAB)+\Tr(DBA) \Bigr)-\Tr(DA)\Tr(DB)
\end{equation*}
  and the variance of a observable is $\Var_{D}(A)=\Cov_{D}(A,A)$.
From the Cauchy--Schwartz inequality we have
\begin{equation*}
\Var_{D}(A)\Var_{D}(B)-\Cov_{D}(A,B)^{2}\geq \frac{1}{4}\vert \Tr(\rho\left[A,B\right])\vert^2
\end{equation*}
  which is known as the Schr\"odinger uncertainty principle \cite{Sch}.
Without the covariance part, one gets the Heisenberg uncertainty relation \cite{Hei}.
The Schr\"odinger uncertainty principle can be reformulated as
\begin{equation*}
\det\begin{pmatrix}
\Cov_{D}(A,A) & \Cov_{D}(A,B)\\
\Cov_{D}(B,A) & \Cov_{D}(B,B)\end{pmatrix}\geq
\det\left\lbrack-\frac{\ci}{2}\begin{pmatrix}
\Tr(D\left[A,A\right]) & \Tr(D\left[A,B\right]) \\
\Tr(D\left[B,A\right]) & \Tr(D\left[A,A\right])\end{pmatrix}\right\rbrack.
\end{equation*}
This form was generalized by Robertson for the set of observables $(A_{i})_{1,\dots,N}$ as
\begin{equation*}
\det\biggl( \left[\Cov_{D}(A_h,A_j)\right]_{h,j=1,\dots,N} \biggr) \geq
\det\biggl( \left[- \frac{\ci}{2}\Tr(D\left[A_h,A_j\right])\right]_{h,j=1,\dots,N} \biggr).
\end{equation*}
In this formula the lower bound is given by the commutators of the observables.

In this paper we prove the inequality
\begin{equation}\label{eq:mainineq1}
\det\biggl( \left[\Cov_{D}(A_h,A_j)\right]_{h,j=1,\dots,N} \biggr) \geq
\det\biggl( \left[\frac{f(0)}{2}\left\langle \ci\left[D,A_{h}\right],\ci\left[D,A_{j}\right]
\right\rangle_{D,f}\right]_{h,j=1,\dots,N} \biggr),
\end{equation}
  where the scalar product $\langle\cdot,\cdot\rangle_{D,f}$ is induced by an operator monotone function
  $f$, according to Petz classification theorem \cite{PetSud}.
The inequality (\ref{eq:mainineq1}) was studied first just in the case $N=1$ for special functions $f$.
The cases $f(x)=f_{SLD}(x)=\frac{1+x}{2}$ and $f(x)=f_{WY}(x)=\frac{1}{4}(\sqrt{x}+1)^{2}$ were proved
  by Luo in \cite{Luo1,Luo2}.
The general case of the Conjecture was proved by Hansen in \cite{Han} and shortly after by Gibilisco,
  Imparato and Isola with a different technique in \cite{GibImpIso1}.

In the case $N=2$ the inequality was proved for $f=f_{WY}$ by Luo, Q. Zhang and Z. Zhang
  \cite{LuoZha2,LuoZha3,LuoZha1}.
The case of Wigner--Yanase--Dyson metric, where
  $f_{\beta}(x)=\frac{\beta(1-\beta)(x-1)^{2}}{(x^{\beta}-1)(x^{1-\beta}-1)}$
  ($\beta\in\left[ -1,2\right]\setminus\left\{0,1\right\}$) was proved independently by Kosaki \cite{Kos}
  and by Yanagi, Furuichi and Kuriyama \cite{YanFurKur}.
The general case is due to Gibilisco, Imparato and Isola \cite{GibImpIso1,GibIso}.
Gibilisco and Isola emphasized the geometric aspects of the inequality (\ref{eq:mainineq1}) and
  conjectured it for general quantum Fisher information \cite{GibIso}.

In a recent paper Gibilisco, Imparato and Isola proved the inequality in the $N=3$ real case
  \cite{GibImpIso2} and conjectured that
\begin{equation}
\label{eq:mainineq2}
\det\biggl( \left[\frac{f(0)}{2}\left\langle \ci\left[D,A_{h}\right],\ci\left[D,A_{j}\right]
  \right\rangle_{D,f}\right]_{h,j=1,\dots,N} \biggr)
\geq
  \det\biggl( \left[\frac{g(0)}{2}\left\langle \ci\left[D,A_{h}\right],\ci\left[D,A_{j}\right]
  \right\rangle_{D,g}\right]_{h,j=1,\dots,N} \biggr)
\end{equation}
  holds too in general if $\frac{f(0)}{f(x)}\geq\frac{g(0)}{g(x)}$.
They proved this conjecture for every $N\in\mathbb{N}$ and for every appropirate function $f$ recently
  in \cite{GibImpIso3}.

In this paper we prove this inequality too in a bit stronger form.

\section{Quantum Fisher information}

The states of an $n$-level system are identified with the set of positive semidefinite self-adjoint  $n\times n$ matrices of trace $1$.
The states form a closed convex set in the space of matrices and denote by $\MN$ its interior, the set of all strictly positive
  self-adjoint matrices of trace $1$.
Let $\TMN$ be the real vector space of all self-adjoint $n\times n$ matrices of trace $0$.

The space $\MN$ can be endowed with a differentiable structure \cite{HiaPetTot}.
The tangent space $T_{D}$ at $D\in\MN$ can be identified with $\TMN$.
A map
\begin{equation*}
K:\MN\times\TMN\times\TMN\to\mathbb{C}\qquad (D,X,Y)\mapsto K_{D}(X,Y)
\end{equation*}
  will be called a Riemannian metric if the following condition hold.
For all $D\in\MN$ the map
\begin{equation*}
K_{D}:\TMN\times\TMN\to\mathbb{C}\qquad (X,Y)\mapsto K_{D}(X,Y)
\end{equation*}
  is a scalar product and for all $X\in\TMN$ the map
\begin{equation*}
K_{\cdot}(X,X):\MN\to\mathbb{C} \qquad D\mapsto K_{D}(X,X)
\end{equation*}
is smooth.

Let $(K^{(m)})_{m\in\mathbb{N}}$ be a family of metrics, such that $K^{(m)}$ is a Riemannian metric on $\Mm$ for all $m$.
This family of metrics defined to be monotone if
\begin{equation*}
K_{T(D)}^{(m)}(T(X),T(X))\leq K_{D}^{(n)}(X,X)
\end{equation*}
  for every stochastic mapping, that is completely positive, trace preserving linear map $T:M_{n}(\mathbb{C})\to M_{m}(\mathbb{C})$,
  for every $D\in\MN$ and for all $X\in\TMN$ and for all $m,n\in\mathbb{N}$.

\begin{theorem}
{\it Petz classification theorem \cite{PetSud}.}
There exists a bijective correspondence between the monotone family of metrics $(K^{(n)})_{n\in\mathbb{N}}$ and operator monotone
  $f:\mathbb{R}^{+}\to\mathbb{R}$ functions such that $f(x)=xf(x^{-1})$ hold for all positive $x$.
The metric is given by
\begin{equation}\label{eq:generatedscalprod}
K_{D}^{(n)}(X,Y)=\Tr\Bigl(X\bigl( R_{n,D}^{\frac{1}{2}} f(L_{n,D}R_{n,D}^{-1}) R_{n,D}^{\frac{1}{2}}\bigr)^{-1}(Y)\Bigr)
\end{equation}
  for all $n\in\mathbb{N}$ where $L_{n,D}(X)=DX$, $R_{n,D}(X)=XD$ for all $D,X\in M_{n}(\mathbb{C})$.
\end{theorem}

For simplicity we denote by $\left\langle\cdot,\cdot\right\rangle_{D,f}$ the scalar product given by Equation (\ref{eq:generatedscalprod})
  in the tangent space of the point $D$.
Let denote by $\Fop$ the set of operator monotone functions $f:\mathbb{R}^{+}\to\mathbb{R}$ with the property $f(x)=xf(x^{-1})$
  for every positive parameter $x$ and with the normalization condition $f(1)=1$.
Here are some elements of the set $\Fop$ from Refs. \cite{Pet1, Pet2}:
\begin{equation*}
\frac{1+x}{2},\ \frac{2x}{1+x},\ \frac{x-1}{\log x},
  \ \frac{2(x-1)^{2}}{(1+x)(\log x)^{2}},
  \ \frac{2(x-1)\sqrt{x}}{(1+x)\log x},
  \ \frac{2x^{\alpha+1/2}}{1+x^{2\alpha}},
  \ \frac{\beta(1-\beta)(x-1)^{2}}{(x^{\beta}-1)(x^{1-\beta}-1)},
\end{equation*}
  where $0\leq\alpha\leq 1/2$ and $0<\vert\beta\vert< 1$.
We also introduce the sets
\begin{equation*}
\Fopr=\left\{ f\in\Fop\ \vert\ f(0)\neq 0\right\}\quad \mbox{and} \quad \Fopn=\left\{ f\in\Fop\ \vert\ f(0)=0\right\}.
\end{equation*}

\begin{theorem} \cite{GibImpIso1}
For $f \in\Fopr$ and $x\in\mathbb{R}^{+}$ set
\begin{equation*}
\tilde{f}(x)=\frac{1}{2}\left((x+1)-(x-1)^2 \frac{f(0)}{f(x)}\right).
\end{equation*}
Then ${\tilde f}\in\Fopn$.
\end{theorem}

For a function $f\in\Fop$ we define
\begin{equation*}
m_{f}(x,y)=xf\left(\frac{y}{x} \right)
\end{equation*}
  which will be used in sequel.

\section{Volume uncertainty principle for Fisher information}

The observables of an $n$-level quantum system are identified with the $n\times n$ self-adjoint matrices, which we will be denoted by $\MT$.
For observables $A,B\in\MT$, state $D\in\MN$ and function $f\in\Fop^{n}$ we define the covariance of $A$ and $B$
  with respect to $D$ and the quantum covariance of $A$ and $B$ with respect to $D$ and $f$ as
\begin{align*}
&\Cov_{D}(A,B)=\frac{1}{2}\Bigl(\Tr(DAB)+\Tr(DBA) \Bigr)-\Tr(DA)\Tr(DB)\\
&\Qov_{D,f}(A,B)=\frac{f(0)}{2}\left\langle\ci\left[D,A\right],\ci\left[D,B\right] \right\rangle_{D,f}.
\end{align*}

\begin{theorem}\cite{GibImpIso1}
Let $\left(\underline{e}_{h}\right)_{h=1,\dots,n}$ be a complete orthonormal base composed of eigenvectors of $D\in\MN$,
  and $\left(\lambda_{h}\right)_{h=1,\dots,n}$ be the corresponding eigenvalues.
For matrices $A,B\in\MT$ we associate matrices $\cpre{A}$ and $\cpre{B}$ whose entries are given respectively by
  $\cpre{A}_{hj}=\langle A_{0}\underline{e}_h,\underline{e}_j\rangle $, $\cpre{B}_{hj}=\langle B_{0}\underline{e}_h,\underline{e}_j\rangle $.
We have the following identities.
\begin{align*}
&\Cov_{D}(A,B)=\Rea\Tr(DA_{0}B_{0})=\frac{1}{2}\sum_{h,j=1}^{n}(\lambda_{h}+\lambda_{j})\Rea(\cpre{A}_{hj}\cpre{B}_{jh}) \\
&\Qov_{D,f}(A,B)=\frac{1}{2}\sum_{h,j=1}^{n}(\lambda_{h}+\lambda_{j})\Rea(\cpre{A}_{hj}\cpre{B}_{jh})
                           -\sum_{h,j=1}^{n}m_{\tilde{f}}(\lambda_{h},\lambda_{j})\Rea(\cpre{A}_{hj}\cpre{B}_{jh}).
\end{align*}
\end{theorem}

In the equations of the previous Theorem the real part function $\Rea$ can be omitted, since the matrices $\cpre{A}$ and $\cpre{B}$ are
  self-adjoint
\begin{equation*}
\cpre{A}_{jh}=\langle A_{0}\underline{e}_h,\underline{e}_j\rangle=\overline{\langle \underline{e}_j,A_{0}\underline{e}_h\rangle}=
  \overline{\langle A_{0}\underline{e}_j,\underline{e}_h\rangle}=\overline{\cpre{A}_{hj}}
\end{equation*}
and for real, symmetric coefficients $\alpha_{h,j}$
\begin{align*}
\sum_{h,j=1}^{n}&\alpha_{hj}\cpre{A}_{hj}\cpre{B}_{jh}=
\sum_{h,j=1}^{n}\alpha_{hj}\Rea(\cpre{A}_{hj}\cpre{B}_{jh})+\ci\sum_{h,j=1}^{n}\alpha_{hj}\Ima(\cpre{A}_{hj}\cpre{B}_{jh})\\
&=\sum_{h,j=1}^{n}\alpha_{hj}\Rea(\cpre{A}_{hj}\cpre{B}_{jh})
  +\ci\sum_{h=1}^{n}\alpha_{hh}\Ima(\cpre{A}_{hh}\cpre{B}_{hh})
  +\ci\sum_{1\leq h<j\leq n}\hskip-5pt\alpha_{hj}\Bigl(\Ima(\cpre{A}_{hj}\cpre{B}_{jh})+\Ima(\cpre{A}_{jh}\cpre{B}_{hj}) \Bigr)\\
&=\sum_{h,j=1}^{n}\alpha_{hj}\Rea(\cpre{A}_{hj}\cpre{B}_{jh})
  +\ci\sum_{1\leq h<j\leq n}\alpha_{hj}\Ima\Bigl(\cpre{A}_{hj}\cpre{B}_{jh}+\overline{\cpre{A}_{hj}\cpre{B}_{jh}} \Bigr)\\
&=\sum_{h,j=1}^{n}\alpha_{hj}\Rea(\cpre{A}_{hj}\cpre{B}_{jh})
\end{align*}
  holds.
So we will use the equations
\begin{align}
&\Cov_{D}(A,B)=\sum_{h,j=1}^{n}\left(\frac{\lambda_{h}+\lambda_{j}}{2}\right)\cpre{A}_{hj}\cpre{B}_{jh} \\
&\Qov_{D,f}(A,B)=\sum_{h,j=1}^{n}\left(\frac{\lambda_{h}+\lambda_{j}}{2}-m_{\tilde{f}}(\lambda_{h},\lambda_{j})\right)\cpre{A}_{hj}\cpre{B}_{jh}
\end{align}
  for covariances.

We will call the set of nonzero matrices $(A_{i})_{i\in I}$ offdiagonally independent if none of them can be written as a linear combination
  of the other matrices and a diagonal matrix.
We call the set $(A_{i})_{i\in I}$ offdiagonally dependent if there are nonzero $(a_{i})_{i\in I}$ coefficients such that the matrix
  $\sum_{i\in I}a_{i}A_{i}$ is diagonal.

\begin{theorem}\label{th:weakv}
Consider a state $D\in\MN$, functions $f,g\in\Fop^{r}$ such that
\begin{equation*}
\frac{f(0)}{f(t)}>\frac{g(0)}{g(t)}
\end{equation*}
  holds and an $N$-tuples of nonzero matrices $(A^{(k)})_{k=1,\dots,N}\in\MT$.
Define the $N\times N$ matrices $\Cov_{D}$ and $\Qov_{D,f}$ with entries
\begin{align*}
&\left[\Cov_{D}\right]_{ij}=\Cov_{D}(A^{(i)},A^{(j)})\\
&\left[\Qov_{D,f}\right]_{ij}=\Qov_{D,f}(A^{(i)},A^{(j)}).
\end{align*}

\item{$(1)$}
We have
\begin{align}
&\det(\Cov_{D})\geq \det(\Qov_{D,f})+\det(\Cov_{D}-\Qov_{D,f})+R(D,f,N),        \label{eq:conj1}\\
&\det(\Qov_{D,f})\geq\det(\Qov_{D,g})+\det(\Qov_{D,f}-\Qov_{D,g})+R(D,f,g,N),   \label{eq:conj2}
\end{align}
where
\begin{align*}
&R(D,f,N)=\sum_{k=1}^{N-1}\binom{N}{k}\bigl\lbrack\det(\Qov_{D,f})\bigr\rbrack^{\frac{k}{N}}
    \bigl\lbrack\det(\Cov_{D}-\Qov_{D,f})\bigr\rbrack^{\frac{N-k}{N}},\\
&R(D,f,g,N)=\sum_{k=1}^{N-1}\binom{N}{k}\bigl\lbrack\det(\Qov_{D,g})\bigr\rbrack^{\frac{k}{N}}
    \bigl\lbrack\det(\Qov_{D,f}-\Qov_{D,g})\bigr\rbrack^{\frac{N-k}{N}}.
\end{align*}

\item{$(2)$}
The following conditions are equivalent.
\begin{align*}
&\mbox{a.}\quad \det(\Cov_{D})=\det(\Qov_{D,f})\\
&\mbox{b.}\quad \det(\Qov_{D,f})=\det(\Qov_{D,g})\\
&\mbox{c.}\quad \mbox{The set of matrices $(A^{(k)}_{0})_{k=1,\dots,N}$ are linearly dependent.}
\end{align*}
\end{theorem}

\begin{proof}
The matrix $\Qov_{D,f}$ is obviously real and symmetric.
First we prove that $\Qov_{D,f}$ is positive definite.
Let us define for indices $1\leq h,j\leq n$
\begin{equation}
\alpha_{hj}^{(f)}=\frac{\lambda_{h}+\lambda_{j}}{2}-m_{\tilde{f}}(\lambda_{h},\lambda_{j})
\quad\mbox{and}\quad
\alpha_{hj}^{(0)}=\frac{\lambda_{h}+\lambda_{j}}{2}.
\end{equation}
Elementary calculations show that $\alpha_{hh}^{(f)}=0$ and for different indices $\alpha_{hj}^{(f)}>0$.
Consider a vector $x\in\mathbb{C}^{N}$ and define an $n\times n$ matrix as $C=\sum_{a=1}^{N}x_{a}\cpre{A}^{(a)}$.
Then we have
\begin{align}\label{eq:posdef}
\left\langle x,\Qov_{D,f}x\right\rangle&=\sum_{a,b=1}^{N}\overline{x_{a}}x_{b}\Qov_{D,f}(A^{(a)},A^{(b)})\\
&=\sum_{a,b=1}^{N}\sum_{h,j=1}^{n}\alpha_{hj}^{(f)}\overline{x_{a}}x_{b}\cpre{A}^{(a)}_{hj}\cpre{A^{(b)}_{jh}}\nonumber\\
&=\sum_{h,j=1}^{n}\alpha_{hj}^{(f)}\overline{\left(\sum_{a=1}^{N}x_{a}\cpre{A}^{(a)}_{jh}\right)}
  \left(\sum_{b=1}^{N}x_{b}\cpre{A}^{(b)}_{jh}\right)\nonumber\\
&=\sum_{h,j=1}^{n}\alpha_{hj}^{(f)}\vert C_{hj}\vert^{2}\geq 0.\nonumber
\end{align}
We note, that if the set of matrices $(\cpre{A}^{(k)})_{k=1,\dots,N}$ are offdiagonally dependent then there exists a nonzero vector
  $x\in\mathbb{C}^{N}$, such that $C=\sum_{a=1}^{N}x_{a}\cpre{A}^{(a)}$ is a diagonal matrix.
For this vector $\left\langle x,\Qov_{D,f}x\right\rangle=0$ holds, and it means that $0$ is an eigenvalue of the matrix $\Qov_{D,f}$, therefore
  its determinant is zero.
If the matrices $(\cpre{A}^{(k)})_{k=1,\dots,N}$ are offdiagonally independent then for every nonzero vector $x$ we have
  $\left\langle x,\Qov_{D,f}x\right\rangle>0$, and in this case $\det(\Qov_{D,f})>0$.
So we have the following equivalence:
\begin{equation}\label{eq:equiv_det_Qovf}
\det(\Qov_{D,f})=0\quad \mbox{if and only if} \quad (\cpre{A}^{(k)})_{k=1,\dots,N}\ \mbox{are offdiagonally dependent.}
\end{equation}

We can repeat our arguments from Equation (\ref{eq:posdef}) for the matrix $\Cov_{D}-\Qov_{D,f}$ using $\alpha^{(0)}-\alpha^{(f)}$
  instead of $\alpha^{(f)}$.
This lead us to the conclusion that $\Cov_{D}-\Qov_{D,f}$ is real, symmetric, positive definite matrix.
Since $\alpha^{(0)}_{hj}-\alpha^{(f)}_{hj}>0$ we have the following equivalence:
\begin{equation}\label{eq:equiv_det_CovQov}
\det(\Cov_{D}-\Qov_{D,f})=0\quad \mbox{if and only if} \quad (A^{(k)}_{0})_{k=1,\dots,N}\ \mbox{are linearly dependent.}
\end{equation}

If for functions $f,g\in\Fopr$
\begin{equation*}\label{eq:equiv_det_QovQov}
\frac{f(0)}{f(t)}>\frac{g(0)}{g(t)}
\end{equation*}
  holds for every positive parameter $t$ then we have $m_{\tilde{f}}(x,y)< m_{\tilde{g}}(x,y)$ for every positive $x$ and $y$.
It means that $\alpha_{hj}^{(f)}-\alpha_{hj}^{(g)}> 0$ for every indices $1\leq h,j\leq n$.
Using the previous arguments we conclude the positivity of the matrix $\Qov_{D,f}-\Qov_{D,g}$ and the equivalence:
\begin{equation}
\det(\Qov_{D,f}-\Qov_{D,g})=0\quad \mbox{if and only if} \quad (A^{(k)}_{0})_{k=1,\dots,N}\ \mbox{are linearly dependent.}
\end{equation}

Using the Minkowski determinant inequality (see for example \cite{BecBel} p. 70.) for real symmetric positive
  matrices $\Qov_{D,f}$ and $(\Cov_{D}-\Qov_{D,f})$ we have
\begin{equation}
\left\lbrack\det( \Qov_{D,f}+(\Cov_{D}-\Qov_{D,f}))  \right\rbrack^{\frac{1}{N}}\geq
\left\lbrack\det( \Qov_{D,f})\right\rbrack^{\frac{1}{N}}+
\left\lbrack\det( \Cov_{D}-\Qov_{D,f} )\right\rbrack^{\frac{1}{N}}
\end{equation}
  and for matrices $\Qov_{D,g}$ and $(\Qov_{D,f}-\Qov_{D,g})$ we have
\begin{equation}
\left\lbrack\det( \Qov_{D,g}+(\Qov_{D,f}-\Qov_{D,g}))  \right\rbrack^{\frac{1}{N}}\geq
\left\lbrack\det( \Qov_{D,g})\right\rbrack^{\frac{1}{N}}+
\left\lbrack\det( \Qov_{D,f}-\Qov_{D,g} )\right\rbrack^{\frac{1}{N}}.
\end{equation}
These equations implies the first part of the Theorem.

To prove the second part of the Theorem assume that $\det(\Cov_{D})=\det(\Qov_{D,f})$.
Using the inequality (\ref{eq:conj1}) we get that $\det(\Cov_{D}-\Qov_{D,f})=0$, which implies that
  $(A^{(k)}_{0})_{k=1,\dots,N}$ are linearly dependent according to the equivalence (\ref{eq:equiv_det_CovQov}).
If $(A^{(k)}_{0})_{k=1,\dots,N}$ are linearly dependent then $\det(\Cov_{D})=0$ and $\det(\Qov_{D,f})=0$, so in this case
  $\det(\Cov_{D})=\det(\Qov_{D,f})$.
Now we proved the equivalence of the $a.$ and $c.$ statements.
The equivalence of $b.$ and $c.$ can be proved similarly.
\end{proof}

\begin{theorem}\label{th:strongv}
Consider a state $D\in\MN$, functions $f,g\in\Fop^{r}$ such that
\begin{equation*}
\frac{f(0)}{f(t)}>\frac{g(0)}{g(t)}
\end{equation*}
  holds and an $N$-tuples of nonzero matrices $(A^{(k)})_{k=1,\dots,N}\in\MT$.
Then for every $t\in\left[0,1 \right]$ parameter the inequalities
\begin{align}
&\det(t\Cov_{D}+(1-2t)\Qov_{D,f})\geq (1-t)^{N}\det(\Qov_{D,f})+t^{N}\det(\Cov_{D}-\Qov_{D,f})+R(D,f,N,t),  \label{eq:conj3}\\
&\det(t\Qov_{D,f}+(1-2t)\Qov_{D,g})\hskip-2pt\geq\hskip-2pt
  (1-t)^{N}\det(\Qov_{D,g})+t^{N}\det(\Qov_{D,f}-\Qov_{D,g})+R(D,f,g,N,t),   \label{eq:conj4}
\end{align}
  hold, where
\begin{align*}
&R(D,f,N,t)=\sum_{k=1}^{N-1}\binom{N}{k}   \Bigl((1-t)\root{N}\of{\det(\Qov_{D,f})} \Bigr)^{k}
                                           \Bigl(    t\root{N}\of{\det(\Cov_{D}-\Qov_{D,f})}\Bigr)^{N-k},\\
&R(D,f,g,N,t)=\sum_{k=1}^{N-1}\binom{N}{k} \Bigl((1-t)\root{N}\of{\det(\Qov_{D,g})} \Bigr)^{k}
                                           \Bigl(    t\root{N}\of{\det(\Qov_{D,f}-\Qov_{D,g})}\Bigr)^{N-k}.
\end{align*}
\end{theorem}
\begin{proof}
The Theorem is just an application of a generalized form of the Minkowski inequality, which is due to Firey \cite{Fir}:
  for $k\times k$ real, symmetric, positive definite matrices $K,L$ and for parameter $t\in\left[0,1 \right]$
\begin{equation*}
\det((1-t)K+tL)^{\frac{1}{k}}\geq (1-t)\det(K)^{\frac{1}{k}}+t\det(L)^{\frac{1}{k}}.
\end{equation*}
\end{proof}

The $t=\frac{1}{2}$ case in the previous Theorem gives back Theorem (\ref{th:weakv}).
\bigskip

{\bf Acknowledgement.}
This work was supported by Japan Society for the Promotion of Science, contract number
  P 06917.

\end{document}